\documentclass[prl,twocolumn,showpacs,amsmath,amssymb,superscriptaddress]{revtex4-1}
\usepackage{graphicx}
\usepackage{ulem}
\usepackage{graphicx}
\usepackage{dcolumn}
\usepackage{bm}
\usepackage{mathrsfs}
\usepackage{subfigure}
\usepackage{natbib}
\usepackage{color}
\usepackage{amssymb}
\usepackage{amsmath}

\newcommand{\etal}{\textit{et al.}}

\begin{document}

\title{Critical behavior at the dynamic Mott transition} 

\author
{Nicola\,Poccia} 
\affiliation{MESA+ Institute for Nanotechnology, University of Twente, 7500 AE Enschede, The Netherlands}
\affiliation{RICMASS Rome International Center for Materials Science Superstripes,
via dei Sabelli 119A, 00185 Roma, Italy}

\author{Tatyana\,I.\,Baturina} 
\affiliation{A.\,V.\,Rzhanov Institute of Semiconductor Physics SB RAS,13 Lavrentjev Avenue, Novosibirsk 630090, Russia}
\affiliation{Novosibirsk State University,
Pirogova str. 2, Novosibirsk 630090, Russia}
\affiliation{Materials Science Division, Argonne National Laboratory,
9700 S. Cass Avenue, Argonne, Illinois 60637, USA}

\author{Francesco\,Coneri}
\affiliation{MESA+ Institute for Nanotechnology, University of Twente, 7500 AE Enschede, The Netherlands}

\author{Cor\,G.\,Molenaar}
\affiliation{MESA+ Institute for Nanotechnology, University of Twente, 7500 AE Enschede, The Netherlands}

\author{X.\,Renshaw\,Wang}
\affiliation{MESA+ Institute for Nanotechnology, University of Twente, 7500 AE Enschede, The Netherlands}

\author{Ginestra\,Bianconi}
\affiliation{School of Mathematical Sciences, Queen Mary University of London, London E1 4NS, United Kingdom}

\author{Alexander\,Brinkman}
\affiliation{MESA+ Institute for Nanotechnology, University of Twente, 7500 AE Enschede, The Netherlands}

\author{Hans\,Hilgenkamp}
\affiliation{MESA+ Institute for Nanotechnology, University of Twente, 7500 AE Enschede, The Netherlands}

\author{Alexander\,A.\,Golubov}
\affiliation{MESA+ Institute for Nanotechnology, University of Twente, 7500 AE Enschede, The Netherlands}
\affiliation{Moscow Institute of Physics and Technology, Dolgoprudnyi, Moscow district, Russia}

\author{Valerii\,M.\,Vinokur}
\affiliation{Materials Science Division, Argonne National Laboratory, 9700 S. Cass Avenue, Argonne, Illinois 60637, USA}

\date{}

\begin{abstract}
We investigate magnetoresistance of a square array of superconducting islands placed on a normal metal, which offers a unique tunable laboratory for realizing and exploring quantum many-body systems and their dynamics. 
A vortex Mott insulator where magnetic field-induced vortices are frozen in the dimples of the egg crate potential by their strong repulsion interaction is discovered.  
We find an insulator-to-metal transition driven by the applied electric current and determine critical exponents that exhibit striking similarity with the common thermodynamic liquid-gas transition.  A simple and straightforward quantum mechanical picture is proposed that describes both tunneling dynamics in the deep insulating state and the observed scaling behavior in the vicinity of the critical point.  Our findings offer a comprehensive description of dynamic Mott critical behavior and establish a deep connection between equilibrium and nonequilibrium phase transitions. 
\end{abstract}

\maketitle

 
 \begin{figure}
\includegraphics[width=1.0\linewidth]{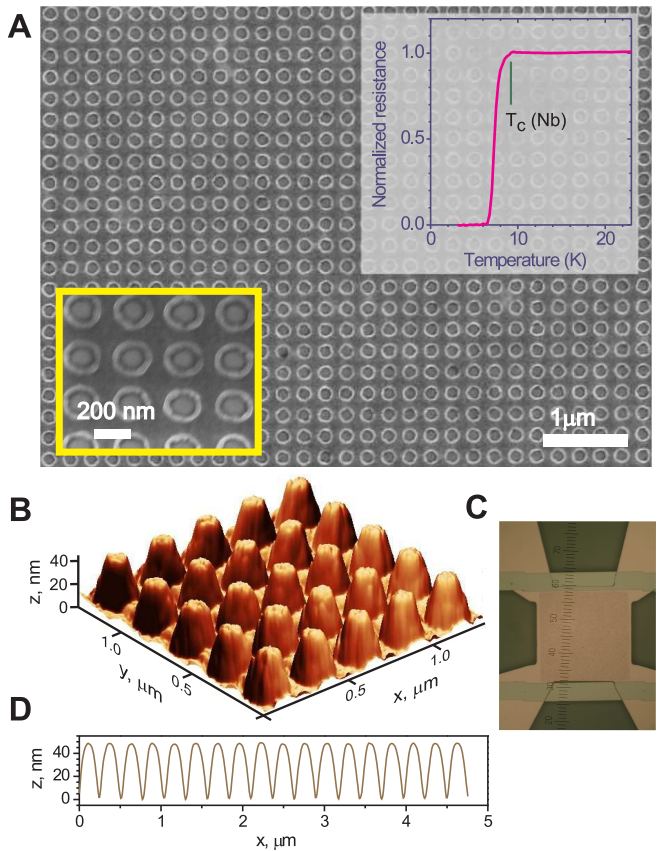}
 \caption{\textbf{Experimental set up.} 
 (\textbf{A}) Scanning electron microscopy (SEM) image of the square array 
 of Nb islands on Au. 
 The bottom-left inset shows a magnified image of the array. 
 The top-right inset shows the temperature dependence of the resistance 
 near the superconducting transition.
 The superconducting critical temperature of bulk Nb is marked by the vertical bar.
 (\textbf{B}) $1.4 \times 1.4\,\mu\textrm{m}^2$ atomic force microscopy topography 
 of the superconductor islands on the metallic template
 of the same sample.
 (\textbf{C}) Optical image of the Au template with four 
 contacts (yellow), Nb bus bars (light green) and the array of $300 \times 300$ Nb islands barely 
 visible as a red square shadow on a Si/SiO$_2$ substrate (dark green). 
 (\textbf{D}) Maximum height profile along the principle axis of the array.}
 \label{fig-main}
 \end{figure}

Mott insulators are systems that are expected to be metallic from the point of view of standard band theory but become insulators owing to correlations in the motions of electrons arising from their Coulomb repulsion. 
If the need of quantum particles to be delocalized to minimize the cost in kinetic energy from spatial confinement prevails over the Coulomb penalty, the system remains metallic.   
Imposing commensurate effects to restrict the freedom of particles to move, one can shift the balance towards localization and cause  transition to insulating 
state~\cite{PM:1937,Mott:1949,Mott:1990,Imada:1998}.
As a practical matter, the transition can be realized by varying pressure or doping, i.e. tuning the carrier density, or by changing either temperature or electric field altering tunneling probability.
The Mott insulator and Mott transition are an exemplary manifestation of
quantum many-body physics of strongly correlated systems and are thought to be
a key ingredient of high temperature superconductivity~\cite{Lee:2006}, quantum phase transitions in a Bose-Einstein condensate and spin liquids~\cite{Jaksch:1998,Hofstetter:2002,Greiner:2002,Bloch:2008,Jordens:2008,Sherson:2010,Weitenberg:2011,Watanabe:2012}, and be fundamental to metal-insulator-semiconductor field-effect transistors and related applications~\cite{Nakano:2012}.
Central to strongly correlated systems are becoming issues related to their \textit{non-equilibrium} behavior, in particular, response to the electric field and the dielectric breakdown of Mott insulators~\cite{Taguchi:2000,Rodichev:2013,Oka:2003,Eckstein:2010} destroying Coulomb localization.  There has been extensive research that strives to describe delocalization and transition to metallic state caused by the electric field based on a many-body Schwinger-Landau-Zener mechanism\cite{Oka:2003,Oka:2010},
nonequilibrium dynamical mean-field theory~\cite{Eckstein:2010}, and time-dependent density-matrix renormalization-group simulations~\cite{Heidrich:2010}. It succeeded to capture the existence of the threshold transition field and produce numerically
a current-voltage dependence at modest fields.
Yet many fundamental questions regarding dynamic Mott transition remain open.
On experimental side, the observed electric field-induced destruction of the Mott
insulator~\cite{Rodichev:2013}, appears as a disorder-promoted 
avalanche-like breakdown along a percolation weak path, rather than a transformation 
of a gapped insulator spectrum into a metallic band-like.
On the theoretical side, the employed approaches are valid at electric fields well below threshold breakdown field, but fail in the close vicinity  
of the transition.  Here we address these challenges. 
We prepared an array of superconducting islands where magnetic field induced vortices are 
localized between the superconducting islands in the areas of weaker proximity-induced
superconductivity, i.e. at the energy dimples of an egg crate potential~\cite{Lobb-Review}.
If thermal fluctuations are not strong enough to overcome the combined localizing action of mutual repulsion 
and egg crate pinning, vortices form a \textit{vortex} Mott insulating state at commensurate fields corresponding to an integer 
number of vortices per pinning site. 
The concept of the vortex Mott state was introduced in~\cite{NelsVin:1993} and addressed in experiments on antidot arrays in superconducting films~\cite{Baert:1995,Harada:1996}. 
Varying magnetic field provides precise control over the vortex number and tunes the ratio of the repulsion between vortices to their mobility~\cite{note1}.  Most importantly, in the structure of choice, it is the period of the superconducting islands array that takes up 
the role of the atomic spacing of an electronic system.  As a result our nanopattern 
harboring vortices
is free of detrimental effects of structural disorder inevitable in electronic systems.
This enables us to observe an unconcealed insulator-to-metal transformation of the 
spectrum.
Adopting a mean-field approximation we derive expressions for 
a dynamic response of a Mott insulator to the applied external drive 
asymptotically exact both at small applied fields and in close vicinity 
of the dynamic transition.  The obtained formulas perfectly describe the experiment and reveal the connection between the equilibrium and nonequilibrium Mott behaviors.


Our samples consist of 40 nm-thick Au, patterned as a four-point set-up
in a Van der Pauw configuration for transport measurements, 
on Si/SiO$_2$ substrates. 
The Au pattern is overlaid 
with a square array of superconducting niobium (Nb) islands 45\,nm thick. 
An array contains 90,000 Nb islands placed with a period $a=267$\,nm. 
The diameter of an island 
is $220\pm 3$\,nm and the island separation is $47\pm 3$\,nm.
Panels A-D of Fig.\,1 show scanning electron, atomic force microscopy, and optical
images of the sample and the height profile along one of the principal axis of the array.
The superconducting transition temperature of the array determined as the
midpoint of the temperature resistance curve in the upper inset in Fig.\,1A, 
is $T_c=7.3$\,K which is 2\,K lower than $T_{c0}=9.3$\,K of bulk niobium. 
This moderate downward shift of the transition temperature implies that the array is 
a strongly coupled network of superconducting islands\cite{Tinkham:1983,Baturina:2005,Mason:2012}. 

\begin{figure}
\resizebox{1.0\columnwidth}{!}{\rotatebox{-00}{\includegraphics{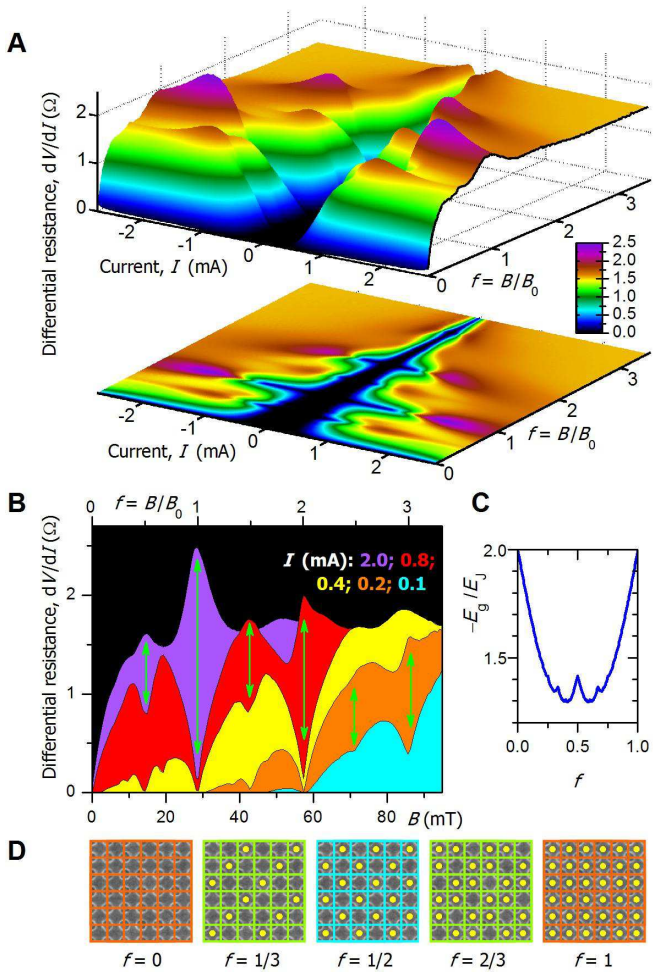}}}
\caption{{\bf Vortex Mott insulator-to-metal transition.} 
(\textbf{A}) Colour plots of the differential resistance as function 
of the applied current and magnetic field in units of $B_0=\pi\hbar/(ea^2)=28.6$\,mT. 
The colour bar gives the differential resistance in Ohms.
(\textbf{B}) Representative differential resistances as functions of the frustration factor $f=B/B_0$ 
taken at different bias currents. 
At low current bias (blue, orange, and yellow), 
the differential resistance minima at 
the frustration factors $f= 1/3, 1/2, 2/3, 1,4/3,3/2,5/3,2,7/3,5/2,8/3,3$ are visible. 
Increasing the current reverses the minima into maxima (red and violet). 
This is a manifestation of the Mott-insulator-to-metal transition and highlighted by vertical arrows 
(the details of the insulator-to metal evolution at $f=1/2,1,$ and 2 are shown below in 
the upper row of Fig.\,3). 
(\textbf{C}) Energy of the ground state, $E_g$, as function of $f$  
obtained by solving the Harper equation~\cite{Harper:1955,Baturina:2011}.
The plot is given for $f\in[0,1]$, since $E_g(f)$ is periodic with the period 1.
(\textbf{D}) Vortex configurations at rational frustration factors $f=0,1/3,1/2,2/3,1$.
The periodic arrays of vortices (yellow circles) superimposed on the 
SEM subimages of the square array of Nb islands on Au shown in Fig.\,1A.
Vortices are trapped in energy minima between superconducting islands.
At integer $f$ the 
vortex Mott insulator forms, 
partial fillings correspond to fractional Mott insulator states.
At integer $f$ the
vortex Mott insulator forms, 
partial fillings correspond to fractional Mott insulator states.  }
\label{3D-plot}
\end{figure}

In order to identify the critical behavior of the dynamic Mott transition we perform
transport measurements in the current-voltage-magnetic field ($I$,$V$,$B$) three-coordinate space
scanning $I$ and $B$ in small steps.
The measurements are carried out in a shielded cryostat at temperature $T=1.4$\,K 
which is much lower than $T_c$.
To probe the density 
of mobile states of the vortex system the differential resistance, $dV/dI$, is used\cite{Deshpande:2009,notezero}.
Panel A of Fig.\,2 shows colour plots of the
differential resistance as function of the applied current and the magnetic field
in units of frustration parameter $f=B/B_0$, where $B_0=\Phi_0/a^2$, and $\Phi_0=\pi\hbar/e$ 
is the magnetic flux quantum.
In our array $B_0=28.6$\,mT.
The data presented in Fig.\,2A enable a meaningful scaling analysis which will 
reveal the nature of the Mott insulator-to-metal transition in the dynamic regime. 
To highlight the transition we display representative isocurrent cuts as $dV/dI$ vs. $B$ plots in 
Fig.\,2B.
At modest currents we reveal a wealth of dips in $dV/dI$ 
at integer frustrations, namely at $f=$1,\,2 and 3\, corresponding to integer number of flux quanta per elemental square of an array,
as well as a fine structure of fractional dips at $f=$1/3,\,1/2,\,2/3,\,4/3,\,3/2,\,...\,. 
These minima reflect the modulation of the ground state energy $E_g$ due to formation of 
periodic vortex patterns in a magnetic field\,(Fig.\,2C,D).
The dips in the resistance and singularities in magnetization at commensurate (but mostly integer) frustrations
due to modulations of $E_g$ were conventionally observed in numerous experiments on Josephson junction-, proximity-, 
and antidot arrays, and superconducting wire networks,
see\cite{Baturina:2011} and references therein.
Importantly, the very observation of minima at fractional $f$ evidences the 
high precision periodicity of our array. Indeed, while the oscillations with the main period,
corresponding to integer $f$, are robust,
the modulations at fractional $f$ require fine synchronization of superconducting phase
in several overlapping contours (see how shallow the corresponding energy minima in Fig.\,2C are) 
and are easily destroyed by even slight irregularities in the array parameters.

 \begin{figure*}
 \begin{center}
 \includegraphics[width=1.0\linewidth]{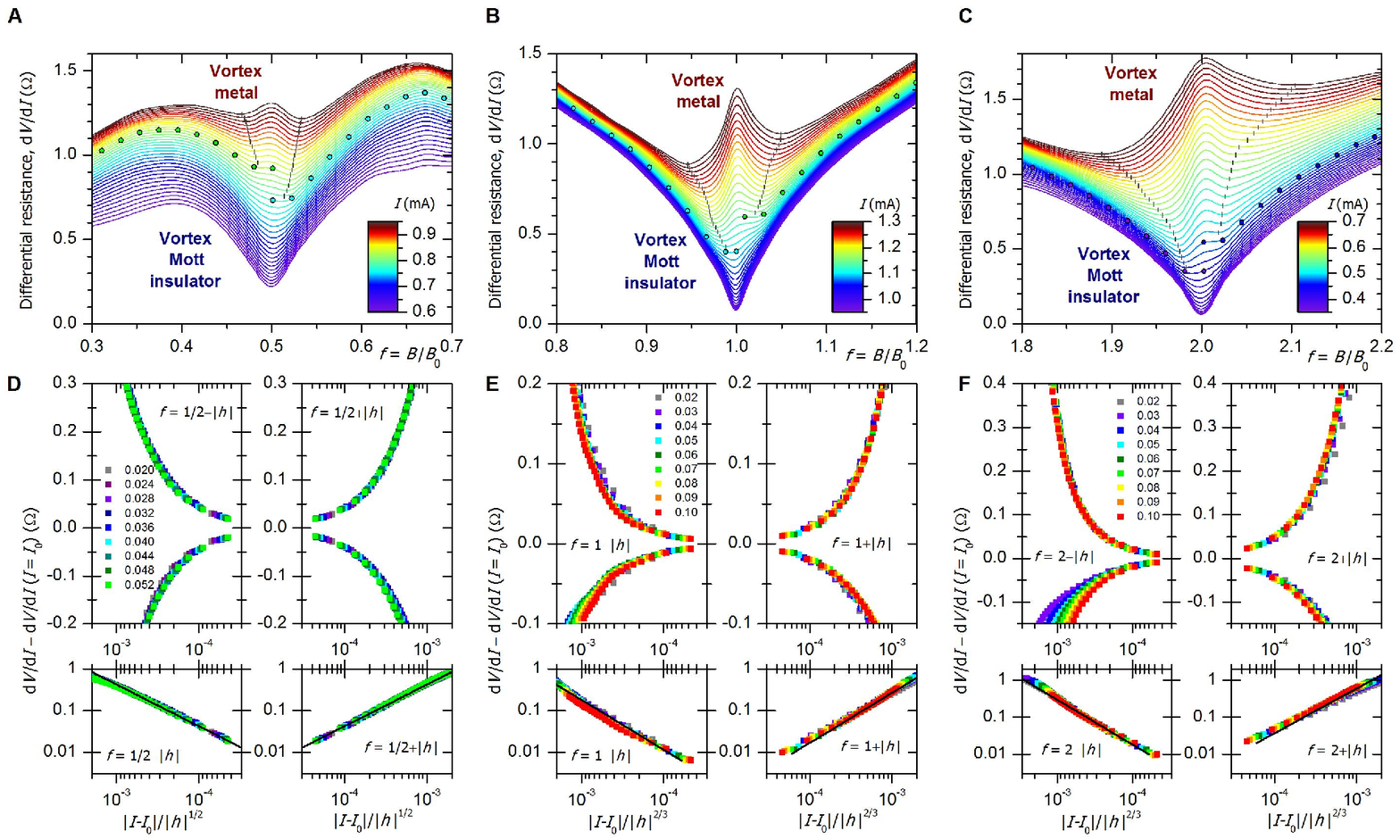}
 \caption{\textbf{Scaling at the dynamic vortex Mott transition}
 (\textbf{A-C}), Differential resistance $dV/dI$ vs. frustration parameter $f$ 
 in the vicinity of $f_c=1/2,1,2$, respectively. 
 Each panel contains 36 traces in 0.01\,mA steps.
 The colours refer to magnitude of the current.
 The dots mark $(dV/dI)(I=I_0)$-s separating insulating-like and metallic-like behaviors.
 The separatrix traces are chosen so that $dV/dI$ start to turn upwards on approach to $f_c$ at $I>I_0$,
 $dV/dI$ maintain downward trend near $f_c$ at $I<I_0$, and $d(dV/dI)/df|_{f=f_c}=0$ at $I=I_0$.
 In the metallic regions $dV/dI$ shows minima marked with strokes which enclose the critical regions
 around $f_c$.
 The traces are not symmetric with respect to $f_c$ so 
 the critical currents for left-, $I_0^{-}$, and right-hand side, $I_0^{+}$, vicinities are different and
 assume the values (\textbf{A}) $I_0^-=0.83$\,mA, $I_0^+=0.78$\,mA; (\textbf{B})
 $I_0^-=1.10$\,mA, $I_0^+=1.14$\,mA; (\textbf{C}) $I_0^-=0.45$\,mA,
 $I_0^+=0.48$\,mA.
 (\textbf{D-F}) Scaling of the same data after subtracting the corresponding separatrix, 
 $dV/dI-(dV/dI)(I=I_0)$ with respect to variable $|I-I_0|/|h|^{(\delta-1)/\delta}$ with 
 $h=f-f_c$ and $\delta=2$ for $f_c=1/2$ (\textbf{D}) and $\delta=3$ for $f_c=1$ and $f_c=2$.
  (\textbf{E,F}) The left-hand side (right-hand side) subpanels present 
  the scaling plots in the $f_c-|h|(+|h|)$ vicinities of the critical frustration parameters. 
 The colors refer to different $|h|$.
 Upper panels employ the linear scales of $dV/dI-(dV/dI)(I=I_0)$, in the lower 
 panels the logarithmic scales are used.
 The scales are the same for each $f_c$ pairwise for left-hand side and right-hand side subpanels.
 The black lines in the lower panels correspond to powers 
 $\mu=1$ for $f_c=1/2$, and to $\mu=1.2\pm 0.03$ for $f_c=1,2$.
 }{\label{fig-scaling}}
 \end{center}
 \end{figure*}

The $E_g(f)$ dependence playing a fundamental role in superconducting regular systems is an energy band edge of the
full energy spectrum [known as Hofstadter's butterfly\cite{Hofstadter:1976}] of the Harper 
equation~\cite{Harper:1955,Rammal:1984,Baturina:2011}.  
This is the same spectrum that arises for tight-binding electrons restricted to a 
two-dimensional square lattice under a magnetic field perpendicular to the plane and which is
the signature of a fractional quantum Hall effect\cite{Dean:2013}.
The identity of the spectra is a manifestation of the duality between vortices and electrical 
charges~\cite{NelsVin:1993,Shahar:1996}.  
Viewing Fig.\,2D as a visualization of Mott-Hubbard model for electrons
the case $f=1$ would correspond to a single particle per site.
Accordingly, for its vortex counterpart, the case of complete filling of the 
pinning sites array ($f_c=1,2, ... $) was conjectured to form a vortex Mott insulator\cite{NelsVin:1993,Nelson:1998}.
Our data enable us to extend this conjecture onto fractional filling and propose a
\textit{fractional Mott insulator}, a mirror image of fractional quantum Hall states.
The profound low current minima in $dV/dI$ at rational $f$ support this conclusion.

Our key observation is a reversal of the minima in $dV/dI$ at rational $f$
into maxima upon increasing the current bias, which, as we show below by scaling analysis, is a direct manifestation of 
the existence of the vortex Mott insulator and its transition into a metallic state. 
We reveal a reach collection of reversals at commensurate fields $f=$1/3, 1/2, 2/3, 1, ... with 
the most pronounced effects being seen at integer and half-integer frustration factors (Fig.\,2B).
The relative strength of the reversal effect follows the respective depths of the $dV/dI$ minima at 
moderate currents.
Similar minima-to-maxima flips themselves were observed in regular superconducting systems of different 
geometries, namely, in
a proximity array~\cite{Benz:1990} and in an antidote array\cite{Jiang:2004}, where models associated 
with vortex depinning were provided. 

The vortex Mott insulating behavior manifests itself
as the tendency to downward divergence of the $dV/dI$ traces on approach to $f_c$.
Remarkably, upon the vortex Mott insulator-to-metal transition with the increasing current, an 
upward divergence of $dV/dI$ traces marks the profound metallic-like $dV/dI$ 
behavior near $f_c$.
Notably, the dip-to-peak evolution upon increasing current is observed only in $dV/dI$,
but not in the $R=V/I$ quantity responsible for the dissipation (see fig.\,S1).
This rules out pure dissipative mechanisms due to depinning as possible origin of the flips 
in $dV/dI$~\cite{Benz:1990,Jiang:2004}.

We present detailed plots of $dV/dI$ vs. $B$ curves in the vicinity of $f_c=1/2, 1, 2$ 
in the critical region in the upper row panels of Fig.\,3. 
There is an asymmetry between the left-hand and right-hand vicinities of the
commensurability points so that they have to be considered independently.
In either of them there is an apparent separatrix between insulating-like and
metallic-like regions. 
We identify the currents $I_0^{\pm}$ corresponding to separatrices between insulating
and metallic behavior as critical currents at which the 
dynamic Mott transition occurs, and find them from the condition $d(dV/dI)/df|_{f=f_c^\pm}=0$
 (asymmetry between $I_0^{\pm}$ may reflect the asymmetry between the `particle' and `hole'
tunneling dynamics near commensurate values of frustration parameter, and parallels the asymmetry between transition temperatures of electrons- and holes-doped Mott insulators~\cite{Segawa:2010}).
These separatrices are highlighted by dots in Fig.\,3A-C.
For $I>I_0^{\pm}$ the respective $dV/dI$ traces display minima, marked by strokes in Fig.\,3A-C
encompassing the region of metallic behavior.
These critical regions widen upon increasing current similar to   
the standard quantum phase transition critical region which increases 
with growing temperature~\cite{Sondhi:1997}.
That $I_0^{-}$ and $I_0^{+}$ are different implies also that in the current intervals $I\in (\min\{I_0^-,I_0^+\},\max\{I_0^-,I_0^+\})$ at exactly commensurate values $f$,  Mott insulating and metallic phases coexist~\cite{Kotliar:1999}.

Now we turn to critical vicinity of the Mott transition critical point, $(I_0,f_c)$.
A scaling theory states that in the critical region of the continuous phase transition 
physical observables follow the scaling laws, so that their respective correlation lengths 
diverge on approach to the critical point~\cite{Kadanoff:1967,Fisher:1974}.
The Mott insulator-to-metal transition is driven by tuning the ratio $U/J$ between 
the repulsion strength $U$ 
and the characteristic tunneling energy (bandwidth), $J$. 
Accordingly, the static Mott transition exhibits scaling behavior 
with respect to temperature, $T$ and the repulsion strength $U$. 
The mean-field version of the Hubbard model predicts~\cite{Kotliar:2000} that near the 
Mott critical point $(T_c,U_c)$, where $T_c$ and $U_c$ are critical temperature 
and critical particle repulsion strength, respectively,
the deviation $\Delta U=|U-U_c|$ should scale as $(T_c-T)^{\delta/(\delta-1)}$, where $\delta$ is the 
exponent relating the scaling of the source field with the order parameter. 
To find the critical behavior of the dynamic \textit{vortex} Mott transition,
we note that the role of $U$ is taken by the magnetic field $B$ which sets the
vortex density and, as such, controls the vortex repulsion. 
Drawing inspiration from the theory of nonequilibrium critical fluctuations
near a dynamic phase transition\cite{Chtchelk:2009} which found that 
the equilibrium singular behavior with respect 
to $T-T_c$ turns into a singular behavior with respect to deviation from the
critical value of applied bias, we conjecture
the following mapping of the critical behavior of the static Mott transition 
to that of dynamic vortex Mott transition: $|U - U_c|\to h = |f_c-f|$,
$|T-T_c|\to|I-I_0|$, and the critical point $(T_c,U_c)\to (I_0,f_c)$ 
yielding critical scaling $|I-I_0|\propto |h|^{(\delta-1)/\delta}$.
To show the validity of this assumption
we scale our $dV/dI$ data into the universal form 
	\begin{equation}
		\frac{dV(f,I)}{dI}-\left[\frac{dV(f,I)}{dI}\right]_{I=I_0}={\cal F_{\pm}}\left(\frac{|I-I_0^{\pm}|}{|h|^{(\delta-1)/\delta}}\right)\,,
	\label{scaling}
	\end{equation} 
taking the exponent $\delta$ as an adjustable fitting parameter. 
The best collapse of the data onto a single curve near $f_c=1/2$ 
is achieved at $\delta=2$ (Fig.\,3D).
Importantly, the independent scaling procedures at both, the left- and right-hand sides of the 
critical frustration yield the same value of $\delta$, despite that separatrices are
different.
The same procedure for the left- and right-hand sides of $f_c=1$ and 2 gives rise
to $\delta=3$ (Fig.\,3E,F). 
Double-logarithmic plots in the vicinity of the above values of $f_c$ display a power-law functional form 
of ${\cal F_{\pm}}(x)\propto x^\mu$, as shown in the lower panels of Fig.\,3D-F. 
For $f_c=1/2$ we find $\mu=1\pm 0.03$. For both $f_c=1$ and $f_c=2$
we find $\mu=1.2\pm 0.03$ for all four plots. 
The revealed universal scaling properties of the current and magnetic field
dependent differential resistance, following Eq.\,1, experimentally
establishes the dynamic vortex Mott transition and the existence of the vortex Mott insulator
in superconducting networks.


To gain insight into the observed scaling behaviors and in how nonequilibrium carriers that are frozen into a Mott insulator start to move in response to external drive, we first consider an \textit{electronic} Mott insulator. The applied electric field stimulates tunneling current $I$ which quantifies the rate
of the decay of the insulating ground state, i.e. dielectric breakdown of the insulator.
The latter occurs via formation of an elemental nucleus of the metallic phase of the
linear size of the order of the electron wavelength $\lambda_{\scriptscriptstyle{\mathrm F}}\simeq a$ and with the characteristic 
energy $E_n$ ($a$ is the atomic spacing of the material). Then the current $I\propto\exp[-E_n/(e{\cal E}a)]$, where $-e$ is an electron charge.  
The elemental volume $a^3$ of an insulator contains $\nu\Delta_ca^3$ electronic states, where $\Delta_c$ is the Mott insulating gap due to Coulomb interactions and $\nu$ is the electronic density of states at the Fermi level. 
The minimal energy that each state should acquire to become metallic, i.e. delocalized, is the energy equal to insulating gap $\Delta_c$.
Thus the nucleation energy $E_{\mathrm n}\simeq\nu\Delta_c^2a^3$.
Remembering that $\nu a^2\simeq 1/v{\scriptscriptstyle{\mathrm F}}$, where $v_{\scriptscriptstyle{\mathrm F}}$ is the Fermi velocity, we find $I\propto\exp(-{\cal E}_{\mathrm{th}}/{\cal E})$, where the threshold breakdown field 
${\cal E}_{\mathrm{th}}=E_n/(ae)\equiv\Delta_c^2/v_{\scriptscriptstyle{\mathrm F}}$. 
Our expression is exactly the decay rate of the ground Mott state obtained by the numerical fit of the dynamics of the one-dimensional Hubbard model~\cite{Oka:2003,Oka:2010,Eckstein:2010,Heidrich:2010} to the
Landau-Zener quantum tunneling formula at ${\cal E}\ll{\cal E}_{\mathrm{th}}$.  
On approach to the threshold field where $ {\cal E}_{\mathrm{th}}-{\cal E}\ll{\cal E}_{\mathrm{th}}$
and the Mott gap vanishes, the barrier for
the electron acquires a parabolic shape with the height proportional to
$ ({\cal E}_{\mathrm{th}}-{\cal E})^{3/2}$~\cite{LO:1986}.
Using the exact formula for the transparency of the parabolic barrier~\cite{LL:1977}, we find the resulting tunneling current as 
$I\propto 1/\{1+\exp[\mathrm{const}(1-{\cal E}_{\mathrm{th}}/{\cal E})^{3/2}]\}$.

To describe a \textit{vortex} Mott insulator we use the duality between vortices and charges~\cite{Fazio:1991,vanWees:1991,NelsVin:1993,Nelson:1998} and replace 
the electric field ${\cal E}$ driving electrons 
by the applied current $I$ dragging vortices. Correspondingly, the
current, which is a response of the electronic system, we substitute by the voltage $V$
induced by moving vortices. 
As a result the $I$-$V$ curves describing the response of a vortex Mott insulator to the applied current read
	\begin{eqnarray}
		&&V\propto\exp(-I_0/I)\,,\,\,\,I\ll I_0;\nonumber \\
		&&V\propto\frac{1}{1+\exp\left[\mathrm{const}\left(1-\frac{I}{I_0}\right)^{3/2}\right]}\,,\,|1-I/I_0|\ll 1\,,
		\label{IV}
	\end{eqnarray}
where $I_0$ is a threshold breakdown current at which a vortex Mott insulator transforms into a metallic phase.  
One sees that the characteristic energy controlling 
the response of the vortex system near the Mott transition scales as $|I-I_0|^{3/2}$, which explains the origin of the experimentally observed scaling $|h|\propto |I-I_0|^{3/2}$
for integer frustrations. Note that near the breakdown threshold the transparency of the tunneling barrier  depends crucially on its shape near the top.  
Since the barriers governing the breakdown at fractional frustrations may differ from those
at integer ones, the derivation of
fractional critical exponents requires special analysis.

 \begin{figure*} [t]
 \centering
 \includegraphics[width=16cm]{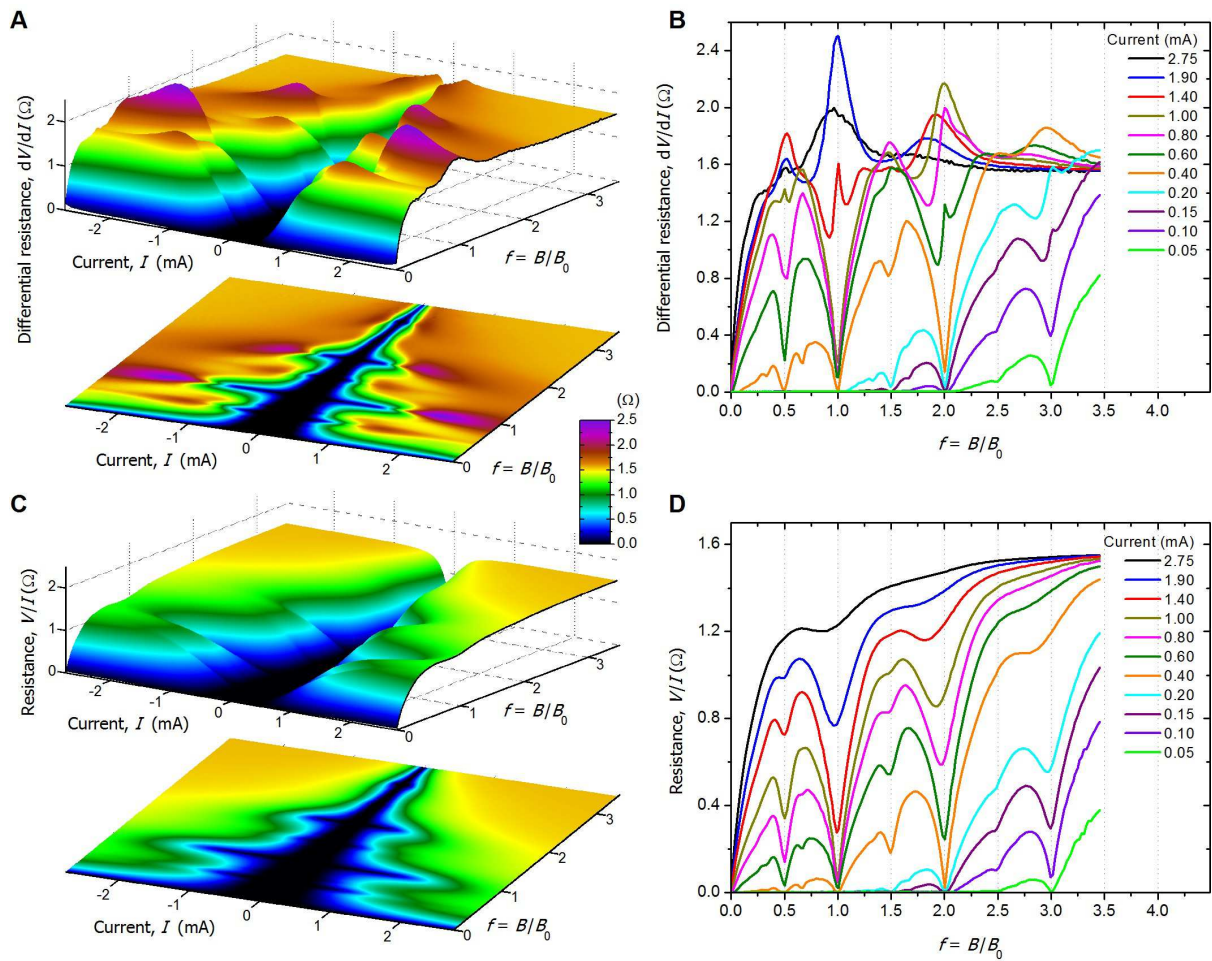}
 \caption {\textbf{(Fig.\,S1) Differential resistance $dV/dI$ vs. resistance $R=V/I$.}
 (\textbf{A}) and (\textbf{C}): color plots of differential resistance and resistance. Color scales are the same 
 for both cases.  (\textbf{B}) and (\textbf{D}): representative curves of differential resistance and resistance as 
 functions of the frustration factor $f=B/B_0$.  Both panels present the curves measured at the same currents, which are denoted by like colors.}
 \label{comaprison}
 \end{figure*}


We now discuss the implications of our results.  
Using an artificially engineered array of superconducting islands to materialize a vortex Mott insulator, enables us to observe the dynamic Mott transition from an insulating to a metallic state as the current-induced transformation of the insulator gapped spectrum into the metallic-like one. 
Notably, our value of $\delta=3$ for the integer \textit{dynamic} vortex  Mott transition coincides with the mean-field value of $\delta$ for the temperature-pressure driven
thermodynamic Mott transition~\cite{Limelette:2003,note}. The latter
was found to belong in the class of the liquid-gas transition of classical systems~\cite{Kadanoff:1967,Kotliar:2000,Limelette:2003} (the insulating phase is associated 
with a ``gas" phase while the metallic phase corresponds to a ``liquid").
The universal scaling properties of the current- and magnetic field-dependent dynamical resistivity experimentally demonstrate that vortex Mott insulator also undergoes a
liquid-gas-like phase transition at the nonequilibrium Mott critical endpoint.
This reveals a deep correspondence
between the dynamic and thermodynamic Mott transitions and supports the earlier
conjecture that the external drive takes
the role of effective temperature in far-from-equilibrium systems~\cite{Chtchelk:2009}.

Quantum mechanical vortex-particle mapping\cite{NelsVin:1993} enables straightforward transcription 
of the results for thermally activated vortex Mott dynamics onto quantum Mott dynamics
of electrons and cold atoms. The mapping is achieved by substitutions $T\leftrightarrow\hbar^{-1}$ and $L\leftrightarrow\hbar/T$, where
$L$ is the size of the system along the magnetic field (i.e. the vortex) direction. This 
makes our system an irreplaceable high-precision laboratory for studying dynamic Mott behaviors in quantum systems
at easily accessible moderate temperatures.

Turning to the difference between critical exponents at integer and fractional frustration parameters 
($\delta=3$ for $f_c=1,2$ and $\delta=2$ for $f_c=1/2$) 
note that an integer vortex Mott insulator corresponds to full filling of the sites.
This state can be viewed as a `ferromagnetic' state and belongs in the Ising universality class.
On the other hand, the case of $f=1/2$ represents an `antiferromagnetic' state since 
the direction of currents in ``empty" cells corresponds to vortices of the opposite direction, fig.\,S2.
Thus, the critical behavior of these two distinct vortex Mott insulating phases 
belongs in different universality classes with different exponents. 
Finally, the energy spectrum of the fractional vortex Mott insulator shown in Fig.\,2C
is the Hofstadter butterfly spectrum of the Harper equation~\cite{Hofstadter:1976},
which is also the spectrum of the fractional quantum Hall electronic state~\cite{Dean:2013}. 
This correspondence between the 
fractional vortex Mott insulating- and fractional quantum Hall electronic states 
offers `vortex-side' evidence for the charge-flux duality previously observed 
near the quantum Hall liquid to insulator transition~\cite{Shahar:1996}.
The gained insights open a route to 
in-depth investigation of dynamic phase transitions in strongly correlated fermionic and bosonic systems using the tunable vortex systems.

\textbf{Acknowledgements:} We thank R. Wiegerink for help and discussion about programming with CleWin software. 
We thank T. Jenneboer for providing the electron beam lithography. 
We thank Gerben Hopman, Pim Reith, Frank Roesthuis and Dick Veldhuis for help 
and support during the experiments. 
N.P. thanks Alessandro Ricci, Gaetano Campi and Antonio Bianconi for valuable discussions. 
The work was supported by the Dutch FOM and NWO foundations, the Russian Academy of Sciences, 
the Russian Foundation for Basic Research (Grant No. 12-02-00152), 
the Ministry of Education and Science of the Russian Federation, and by the U.S. Department of Energy, 
Office of Science, Materials Sciences and Engineering Division. 
N.P acknowledges for financial support the Marie Curie IEF project FP7-PEOPLE-2012-IEF-327711-IMAX.

\section*{Supporting Materials}

\begin{figure}[b]
\centering
\resizebox{1.0\columnwidth}{!}{\rotatebox{-00}{\includegraphics{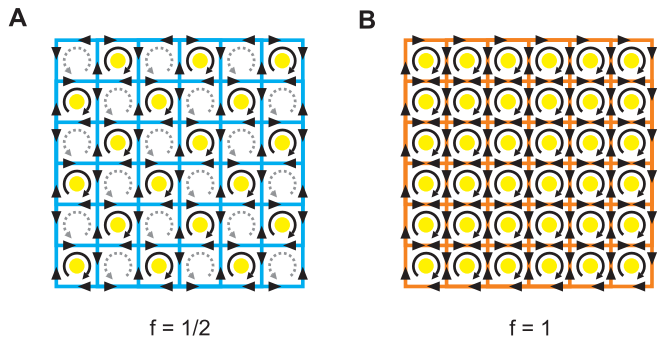}}}
\caption{\textbf{(Fig.\,S2) Current configurations at half-filled, $f=1/2$, and filled, $f=1$, states.}
(\textbf{A}) $f=1/2$: Vortices, shown as yellow circles, fill every other site in the alternating checker-board order and induce 
clockwise currents, shown as solid round arrows, in the filled plaquettes. This implies that every contour 
corresponding to an empty plaquette carries an anti-clockwise current (dotted arrows), as if it contained a
vortex of the opposite direction.  Therefore, such a state can be referred to as a ``antiferromagnetic" state.
(\textbf{B}) $f=1$: every site contains a vortex inducing clockwise currents.  Therefore currents passing through every internal bond
cancel each other and the only nonzero current is the current circumventing the external contour of the system.
Thus with respect to the internal current distribution, the state with $f=1$, the ``ferromagnetic" state is equivalent to the 
state with $f=0$, where trapped vortices are absent.  }
\label{RBosc}
\end{figure}

\subsection*{Methods}
Standard photolithographic tools have been employed to create a 40\,nm thick Au template 
(see Fig.\,1) on a Si/SO${_2}$ substrate of 1$\times$\,1\,cm$^2$. 
The template consists of a central square of 82$\times$\,82\,$\mu$m$^2$, 
with the corners connected to 4 
terminals for electric measurements (not visible in figure). 
The size of the terminals is 200\,$\times$\,200\,$\mu$m$^2$, large enough for micro bonding. 
The Nb pattern is then created on top of the central Au square employing electron beam lithography 
and DC sputtering, via a standard PMMA lift off procedure. 
In this way the electrical conduction among the Nb islands can take place via the Au template 
that lies underneath. 
The resulting Nb array is 80$\times$\,80\,$\mu$m$^2$ large and comprises 300\,$\times$\,300 sites. 
The Nb island on each site has a diameter of $r=220 \pm3 $\,nm and is $45 \pm 3$ nm thick. 
The average distance between the islands is $47 \pm 3$\,nm.
We performed 4-point probe resistivity measurements to determine 
the superconducting transition temperature. 
To ensure a uniform current injection into the array, two 45\,nm Nb bus bars are deposited along 
two opposite sides of the array, as shown in Fig.\,1C; the gap between strips and array 
is less that 0.5\,$\mu$m. 
When cooled below 9\,K, the bars become superconducting and effectively short 
the corresponding current and voltage leads. 
This results in a 2-point measurement that yields 
magnetoresistance data shown in Fig.\,2.
The $IV$ measurements are carried out in a shielded cryostat at 1.4\,K. 
A current bias is applied using a ramp generator at several Hz. 
The measured voltage waveform is amplified using homebuilt low-noise electronics, 
and subsequently digitized with a National Instruments PXI-4461. 
The $IV$ characteristics are non-hysteretic and averaged over several current cycles. 
The dynamic resistance is found by numerically differentiating the averaged $IV$ characteristics. 
A magnetic field is applied by placing a solenoid around the sample with the field perpendicular 
to the sample plane. 
The current through the solenoid increases in a stepwise fashion and separate $IV$ traces 
are recorded at each field step. 
The obtained $IV$ curves are nonlinear, similar to superconducting junction characteristics 
but with a residual resistance. 
This resistance corresponds to the finite distance between the superconducting bus bars 
and the Nb island array.

\subsection*{Differential resistance $dV/dI$ and resistance $R=V/I$}
Figure S1 shows the same set of data as in the main text, Fig.\,2A, presented as differential resistance, panels (A) and (B), and 
as a resistance, panels (C) and (D). All the measurements were performed under the dc applied 
current. Several important comments are in order: 

(i) The $dV/dI$ reaches much larger values than the 
corresponding values of the resistance, as clearly seen in the color plots.

(ii) Dip-to-peak evolution at rational $f$ upon the increasing current is experienced only by
$dV/dI$ but not by the resistance.

(iii) As the measurements are carried out under the constant currents, the plots for the resistances 
coincide up to the numerical factor, $I^2$, with the plots for the dissipated power. 
Upon increasing current the dissipation grows, as it should, since small pinning barriers 
are washed up by larger drives.  At the same time the magnitude of the dissipation 
still exhibit local minima at rational $f$ as is clearly demonstrated by almost all curves
at panel (D).

\subsection*{`Ferromagnetic' and `antiferromagnetic states'}
Figure S\,2 shows current distribution in the proximity network corresponding to the half-integer, $f=1/2$, 
and integer, $f=1$, filling.  At $f=1/2$ the distribution of currents correspond to alternating vortices 
of opposite direction and therefore is referred to as an ``antifferromagnetic" state.  Accordingly,
the state with $f=1$ can be viewed as the ``ferromagnetic" one.  The current distributions at 
and the ground states for vortices trapped in the egg-crate potential relief are described in detail in a review by
Newrock\cite{Lobb-Review} 
Since antiferromagnetic and ferromagnetic states belong in different universality classes
one expects  the scaling exponents $\delta$ characterizing the behaviour of the order parameter to
differ.






%





\begin{thebibliography}{1}
\bibitem{note1}
The vortex density and, therefore the strength of their interaction is proportional 
to the magnetic field. At the same time magnetic field suppresses Josephson coupling between
the islands and thus diminishes the tunneling barrier.
\end{thebibliography}
\end{document}